# Research Data Explored II: the Anatomy and Reception of figshare[1]

Peter Kraker[*], Elisabeth Lex[**], Juan Gorraiz[***], Christian Gumpenberger[***] and Isabella Peters[****]

[*]*pkraker@know-center.at*
Know-Center, Inffeldgasse 13/VI, Graz, 8010 (Austria)

[**]*elisabeth.lex@tugraz.at*
Graz University of Technology, Institute for Knowledge Technologies, Inffeldgasse 13/V, Graz, 8010 (Austria)

[***]*christian.gumpenberger | juan.gorraiz@univie.ac.at*
University of Vienna, Vienna University Library, Dept of Bibliometrics, Boltzmanngasse 5, Vienna, 1090 (Austria)

[****]*i.peters@zbw.eu*
ZBW Leibniz Information Centre for Economics, Düsternbrooker Weg 120, Kiel, 24105 & Kiel University, Christian-Albrechts-Platz 4, Kiel, 24118 (Germany)

**Introduction**

We are currently witnessing a change in scholarly communication. Next to the paper, complementary materials, such as research data, source code, and images are regarded as important outcomes that should be shared and built upon (Kraker et al., 2011). In this new ecosystem, many archives have been established that cater to the needs of a digital and open science. With the increasing importance of research data in the last years, these archives are now receiving initial interest from bibliometrics research.

Torres-Salinas et al. (2013 & 2014) have performed first coverage and citation analyses of research data in Data Citation Index (DCI). Their results have been corroborated by a further study performed by the authors of this paper (Peters et al., 2015). We found that while research data remain mostly uncited (about 85%), there has been a growing trend in citing data sets published since 2008. We have also studied the frequency of altmetrics scores for cited research data. The results show that the number of cited research data with altmetrics "foot-prints" is even lower (4 to 9%) but hint at a higher coverage of research data from the last decade. However, no relationship between the number of citations and the total number of altmetrics scores could be observed. Certain data types (i.e. survey, aggregate data, and sequence data) are more often cited and do receive higher altmetrics scores, but results vary depending on the research field. One of the more surprising results of our first study was that none of the items from figshare, which is one of the largest multidisciplinary repositories for research materials to date, has received more than one citation in the DCI.

In this paper, we investigate figshare (http://figshare.com) more deeply. For this purpose, we analysed the structure of items archived in figshare, their usage, and their reception in two altmetrics sources with a focus on datasets and filesets. Specifically, we addressed the following questions:

[1] The Know-Center is funded within the Austrian COMET program – Competence Centers for Excellent Technologies - under the auspices of the Austrian Federal Ministry of Transport, Innovation and Technology, the Austrian Federal Ministry of Economy, Family and Youth, and the State of Styria. COMET is managed by the Austrian Research Promotion Agency FFG.

- How are document types distributed in figshare? How have different types developed over time? Who are the main providers of items in figshare?
- How are usage data distributed in figshare? How are they correlated?
- To what extent are figshare items visible on various altmetrics channels? Do results from providers of altmetric scores (e.g., PlumX) differ?

**Data source**

Three different data sources were used in this study: (i) figshare, (ii) PlumX and (iii) ImpactStory. figshare is a multidisciplinary repository for research materials that was founded by Mark Hahnel in 2011; it has subsequently been supported by Digital Science of Macmillan Publishers. figshare offers a limited amount of free storage space for private use, and an unlimited amount of storage space for publicly shared materials. Currently, users may upload figures, media (videos, audios), posters, papers, theses, code (source code and binaries), presentations, datasets, and filesets. According to figshare datasets "usually provide raw data for analysis. This raw data often comes in spreadsheet form, but can be any collection of data, on which analysis can be performed." Typical file formats include CSV, XLS, XLSX and SAV. Filesets, on the other hand, "offer a solution to those wishing to group multiple files as a single citable object. These are often experiments or workflows." (figshare n.d.a)

Each item on figshare can be assigned to multiple sub-disciplines that are grouped in 13 main disciplines. As soon as a user makes an uploaded material publicly available, it gets allocated a DataCite DOI. figshare also keeps track of usage data, and displays selected statistics on the page of each item (figshare, n.d.b). In addition, figshare hosts the supplemental data for all PLOS journals (figshare, 2013).

Altmetric scores are spread over a variety of social media channels, e.g., Twitter, Mendeley, or Facebook, which often enough results in complicated and time-consuming approaches for data collection. PlumX and ImpactStory are among the most popular collectors of social media data, providing convenient, but fee-based access to altmetric scores. PlumX targets institutions (e.g., publishers, libraries, or universities) whereas ImpactStory works best for individual researchers. The aggregators differ in the number of social media channels and type of permanent identifiers (e.g., DOI, PubMedID) searched as well as the output format (e.g. json) of retrieved altmetric scores (Chamberlain, 2013). Inevitably, the aggregators' attributes will result in different numbers of documents as well as altmetric scores found (Chamberlain, 2013; Zahedi, Fenner, & Costas, 2014). Jobmann et al. (2014) showed that Plum Analytics had highest coverage and scores in Mendeley and Facebook, whereas ImpactStory recorded a higher coverage of Twitter.

**Methodology**

We used the figshare API to retrieve the basic metadata for all publicly available records up until (excluding) December 2, 2014. We retrieved the metadata for 1,092,808 items. The following fields were used in the analysis: defined_type, published_date, DOI. We then gathered extended metadata for all datasets and filesets with a unique DOI (n=266,961 items). From this dataset, we used the following fields for the analysis: categories, downloads, views[2]. Both datasets are openly available (see Kraker et al., 2015).

---

[2] We did not investigate shares, due to the unclear semantics of this term in figshare."shares" does not refer to actual shares on social media platforms; it is rather counted how often one of the sharing buttons is clicked on the page of the individual item.

Subsequently, the top 500 downloaded items and the top 500 viewed items were analysed with PlumX and ImpactStory via their DOIs. Their coverage on social media platforms and the altmetric scores were compared. The analyses in PlumX were performed at the end of January 2015; the ImpactStory download took place between December 15[th], 2014 and January 27[th], 2015.

**Results**

*General Analysis of figshare*
Tables 1 and 2 show the distribution of basic metadata on figshare. Table 1 reveals that most content in figshare is provided by PLOS (89.2%). The PLOS journal with most items is PLOS ONE with 829,243 items, representing ¾ of the materials archived by figshare (75.9%).

Table 1. DOI providers (n=1,092,808 items)

| DOI providers | items |
|---|---|
| PLOS ONE | 829,243 (75.9%) |
| figshare | 117,572 (10.8%) |
| PLOS Genetics | 36,775 (3.4%) |
| PLOS Pathogens | 33,245 (3.0%) |
| PLOS Computational Biology | 29,517 (2.7%) |
| PLOS Neglected Tropical Diseases | 20,376 (1.9%) |
| PLOS Biology | 17,141 (1.6%) |
| PLOS Medicine | 8,798 (0.8%) |
| none | 141 (0.0%) |

Table 2 presents the basic distribution of material type on figshare. Figure is the type with most items, followed by dataset and filesets. We found that PLOS has the largest share of materials in filesets (92.2%).

Table 2. Document types in figshare with a DOI (n=1,092,808 items)

| type | #items | figshare | PLOS | none |
|---|---|---|---|---|
| figure | 747,207 | 71,400 (9.6%) | 675,799 (90.4%) | 8 |
| dataset | 261,721 | 22,484 (8.6%) | 239,177 (91.4%) | 60 |
| fileset | 64,776 | 5,027 (7.8%) | 59,748 (92.2%) | 1 |
| paper | 14,369 | 14,351 (99.9%) | 18 (0.1%) | 0 |
| presentation | 1,434 | 1,434 (100%) | 0 | 0 |
| poster | 1,429 | 1,429 (100%) | 0 | 0 |
| media | 1,313 | 888 (67.6%) | 353 | 72 |
| code | 366 | 366 (100%) | 0 | 0 |
| thesis | 193 | 193 (100%) | 0 | 0 |

Figure 1 shows that the number of research products is steadily increasing across all document types (except for "figure" where a significant drop in number can be noticed for 2012).

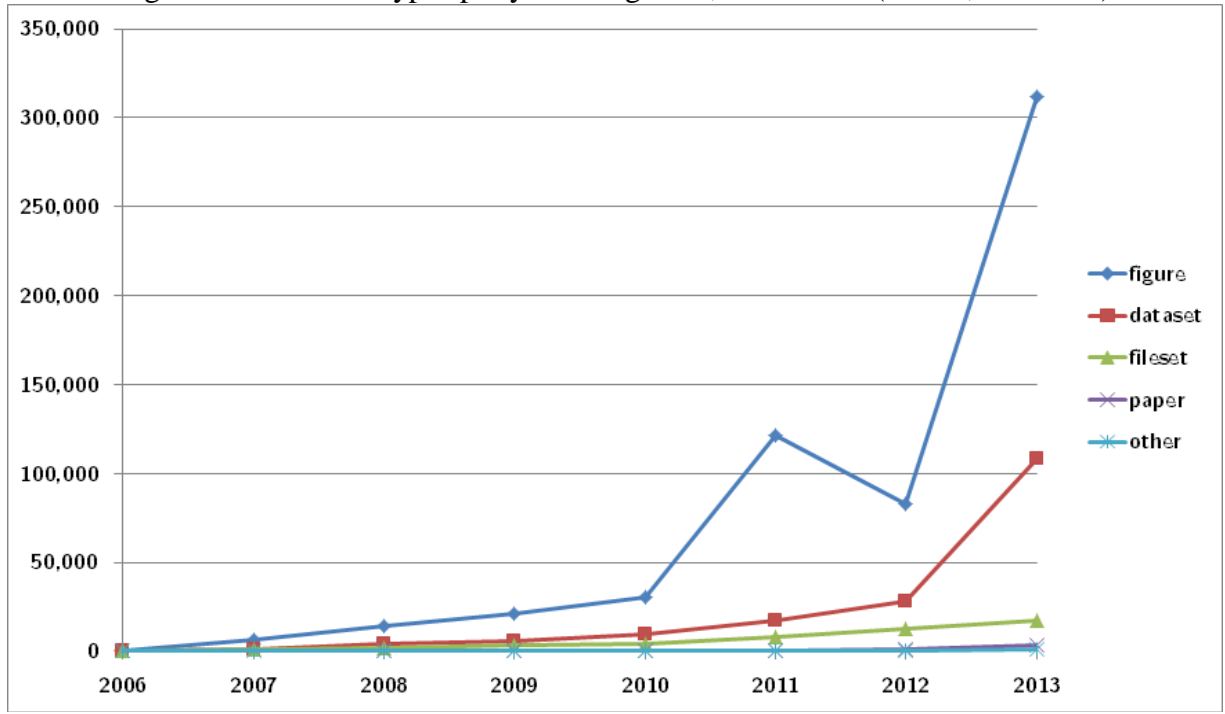

Figure 1. Document types per year in figshare; 2006-2013 (n=818,108 items)

*Usage Analyses for datasets and filesets*
Figure 2 shows the distribution of view and download frequencies among datasets and filesets. Both distributions are highly skewed, exhibiting that only a small fraction of items in figshare are highly used.

We also investigated whether views and downloads are correlated. Since the data is not normal distributed, we computed the Spearman correlation, which resulted in a correlation coefficient of 0.28. A more in-depth analysis of the distribution of the data revealed that out of 266,961 entries, 102,148 have 0 views and 0 downloads. We performed a spot-check investigation to analyse from which providers these entries come from. We found that many entries that have no views or downloads come directly from PLOS whereas entries that have a large number of views (>= 100 Views, 2 entries even have > 10,000 views) are posted directly on figshare. We also found that downloads and views follow a power-law distribution. Consequently, views and downloads follow the principle of preferential attachment which means that items that have many views/downloads will be more likely viewed/downloaded.

Table 3 shows the distribution of disciplines among filesets and datasets. 88.9% of all items have been assigned to Biological Sciences, which makes it the top discipline. Chemistry comes second and Earth and Environmental Sciences third. When leaving out items from PLOS, Biological Sciences is still the largest discipline, but Engineering and Social Science become the second and third largest discipline respectively.

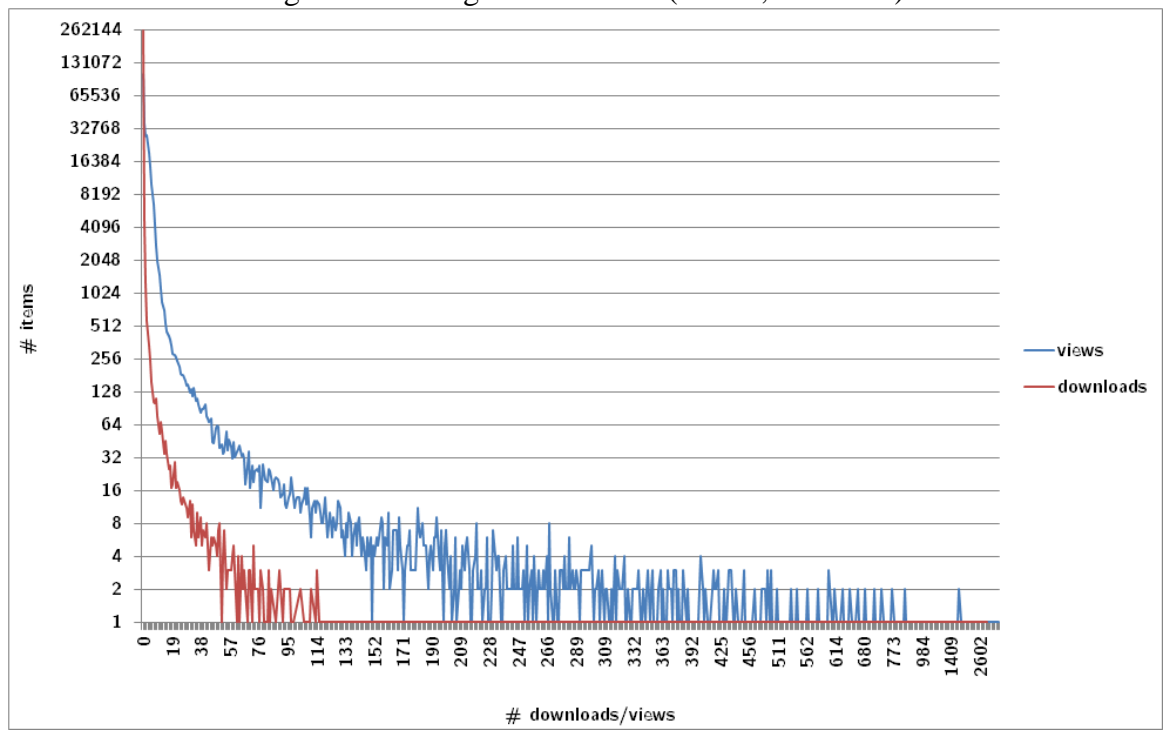

Figure 2. Distribution of view and download frequencies among datasets and filesets in figshare on a logarithmic scale (n=266,961 items)

Table 3. Distribution of disciplines among filesets and datasets, multiple choices possible (n=266,961 items)

| discipline | #items | figshare | PLOS | none |
|---|---|---|---|---|
| Biological Sciences | 237,211 | 8,665 | 228,561 | 58 |
| Chemistry | 39,535 | 3,764 | 35,789 | 3 |
| Earth and Environmental Sciences | 23,260 | 2,840 | 20,417 | 4 |
| Meta Science (e.g., science policy and survey results) | 13,660 | 407 | 13,236 | 17 |
| Mathematics | 12,417 | 362 | 12,056 | 2 |
| Information And Computing Sciences | 11,130 | 1,761 | 9,374 | 1 |
| Social Science | 9,814 | 4,138 | 5,676 | |
| Engineering | 8,448 | 6,217 | 2,231 | |
| Psychology | 7,195 | 306 | 6,892 | |
| Physics | 5,800 | 521 | 5,288 | 1 |
| Uncategorised | 4,218 | 284 | 3,933 | 1 |
| Humanities | 1,665 | 1,665 | | |
| Astronomy, Astrophysics, Space Science | 1,248 | 1,221 | 27 | |
| Health Sciences | 521 | 521 | | |

*Results of the altmetrics analysis*
As mentioned above, we performed an altmetrics analysis in PlumX and ImpactStory of the top 500 downloaded and top 500 viewed datasets and filesets. 243 items appeared in both samples; that means almost 50% of the most viewed items were also most downloaded. Thus, we analysed all unique items (i.e., DOIs) in both samples (n=757). We first report the results of the individual analyses and then compare the outcome of both aggregators with respect to coverage and scores.

*Results of PlumX*

Table 4. General results of the analysis in PlumX. DT=document type (n=757 items)

| Type | Data | File Set | Other | All DTs |
|---|---|---|---|---|
| # items total | 281 | 456 | 20 | 757 |
| # items with captures | 8 | 14 | 1 | 23 |
| Captures Total | 9 | 36 | 18 | 63 |
| Captures Mean | 0.03 | 0.08 | 0.90 | 0.08 |
| Captures Maximum | 2 | 12 | 18 | 18 |
| # items with social media | 214 | 294 | 16 | 524 |
| Social media Total | 4313 | 6100 | 448 | 10861 |
| Social media Mean | 15.35 | 13.38 | 22.40 | 14.35 |
| Social media Maximum | 635 | 388 | 115 | 635 |
| #items with mentions | 17 | 23 | 1 | 41 |
| Mentions Total | 95 | 89 | 1 | 185 |
| Mentions Mean | 0.34 | 0.20 | 0.05 | 0.24 |
| Mentions Maximum | 24 | 22 | 1 | 24 |
| # items with usage/views | 281 | 456 | 16 | 753 |
| Usage - Views - Total | 138717 | 236829 | 11987 | 387533 |
| Views Mean | 493.65 | 519.36 | 599.35 | 511.93 |
| Views Maximum | 4992 | 26985 | 5764 | 26985 |
| # items with usage/downloads | 242 | 445 | 16 | 703 |
| Usage - Downloads - Total | 7296 | 30292 | 1863 | 39451 |
| Downloads Mean | 25.96 | 66.43 | 93.15 | 52.11 |
| Downloads Maximum | 582 | 4534 | 1240 | 4534 |
| # items with scores | 281 | 456 | 16 | 753 |
| **Total Scores** | 150430 | 273346 | 14317 | 438093 |
| **Scores Mean** | 535.34 | 599.44 | 715.85 | 578.72 |
| **ScoresMaximum** | 5020 | 30306 | 7005 | 30306 |

Table 4 shows general results of the analysis in PlumX. The results exhibit a low number of scores in the categories "captures", "mentions" and "social media". In contrast to that, usage numbers were a lot higher. Interestingly, PlumX did not identify all items as "data" or "file sets" as was expected, but created further document types (the category "other" comprises of 3 articles, 1 code, 2 figures, 8 papers, 4 posters, and 1 presentation, accounting for 2.6% of the sample).

Tables 5, 6 and 7 show the origin of the scores in PlumX for captures and mentions, social media and usage. Table 5 indicates that almost all of the captures originate from Mendeley, whereas mentions are predominantly comments in Facebook[3].

Table 5. Captures and mentions in PlumX for each document type

| Document Type | Captures | | Mentions | | |
|---|---|---|---|---|---|
| | Bookmarks: Delicious | Readers: Mendeley | Comments: Reddit | Comments: Facebook | Links Wikipedia |
| Data | 1 | 8 | 6 | 84 | 5 |
| File Set | | 36 | 1 | 88 | |
| Other (Paper) | | 18 | | 1 | |
| Total | 1 | 62 | 7 | 173 | 5 |
| % | 1.59% | 98.41% | 3.78% | 93.51% | 2.70% |

Table 6 suggests that Twitter is the predominant provider of altmetric scores, followed by Google+ –besides the number of shares in figshare, as already reported above.

Table 6. Social Media in PlumX for each document type

| Document Type | Social Media | | | | | |
|---|---|---|---|---|---|---|
| | Scores: Reddit | Shares: Figshare | Shares: Facebook | +1s: Google+ | Tweets: Twitter | Likes: Facebook |
| Data | 26 | 1214 | 224 | 448 | 2245 | 156 |
| File Set | 2 | 2122 | 349 | 692 | 2638 | 297 |
| Other | 0 | 82 | 17 | 48 | 293 | 8 |
| Total | 28 | 3418 | 590 | 1188 | 5176 | 461 |
| % | 0.26% | 31.47% | 5.43% | 10.94% | 47.66% | 4.24% |

Table 7 shows that almost all the usage data for the document type "data" are coming from figshare. For the document type "file sets", we also retrieved scores in EBSCO, PLOS and PubMed, but in very low proportion (less than 10% of the downloads, or 1% of the total usage volume). Please note that some subcategories (like PDF-Views or HTML-Views, see Table 7) could be assigned to both groups. In our analysis, they were considered as "downloads" rather than "views". Due to their low amount number, however, they should not distort the general results.

---

[3] Comments in Reddit and in Facebook are assigned to the category "mentions" in PlumX, although they could equally be assigned to "Social Media". In our comparison of PlumX and ImpactStory, comments, likes and shares in Facebook are aggregated to one group or type.

Table 7. Usage data in PlumX for each document type

| Document Type | Usage | | | | | | | | |
|---|---|---|---|---|---|---|---|---|---|
| | Views | | | Downloads | | | | | |
| | Views: Figshare | Abstract Views: EBSCO | Clicks: Bitly | PDF Views: PubMed Central | PDF Views: PLoS | PDF Views: EBSCO | HTML Views: PubMed Central | HTML Views: PLoS | Downloads: Figshare |
| Data | 138245 | | 472 | | | | | | 7296 |
| File Set | 236421 | 33 | 375 | 16 | 760 | 13 | 67 | 3666 | 25770 |
| Other | 2381 | 0 | 247 | 0 | 0 | 0 | 0 | 0 | 206 |
| Total | 386372 | 33 | 1128 | 16 | 760 | 13 | 67 | 3666 | 34929 |
| % Views/Downloads | 99.70% | 0.01% | 0.29% | 0.04% | 1.93% | 0.03% | 0.17% | 9.29% | 88.54% |
| % Usage | 90.49% | 0.01% | 0.26% | 0.00% | 0.18% | 0.00% | 0.02% | 0.86% | 8.18% |

Finally, Figure 3 shows the mean values for each category (captures, mentions, social media and usage data) depending on the publication year of the document. The mean values of all groups have strongly increased since 2012 despite of the shorter analysis window.

Figure 3. Mean altmetric scores for each publication year from PlumX results

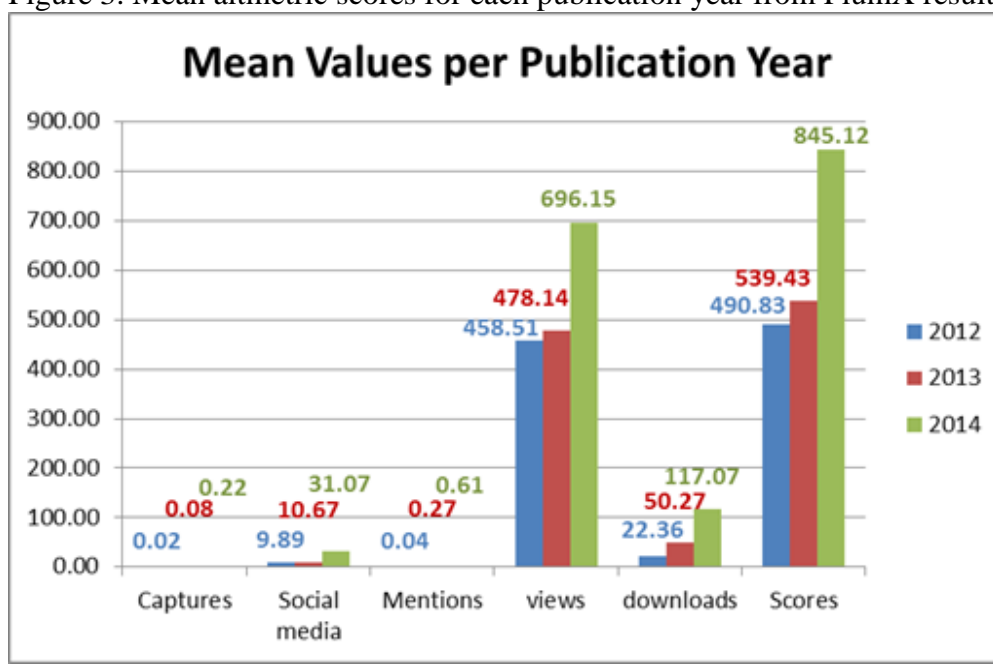

*Results of ImpactStory*

For the 757 unique DOIs, ImpactStory reported altmetrics scores for 455 DOIs (60.1%). Figure 4 shows that items from 2014 attracted the most attention across social media services. The highest altmetrics scores per year are found for tweets followed by blog posts. The increasing number of blog posts is also crucial for the good reception of figshare data in 2014.

Figure 4. Mean altmetric scores for each publication year from ImpactStory results[4] (n=455 items)

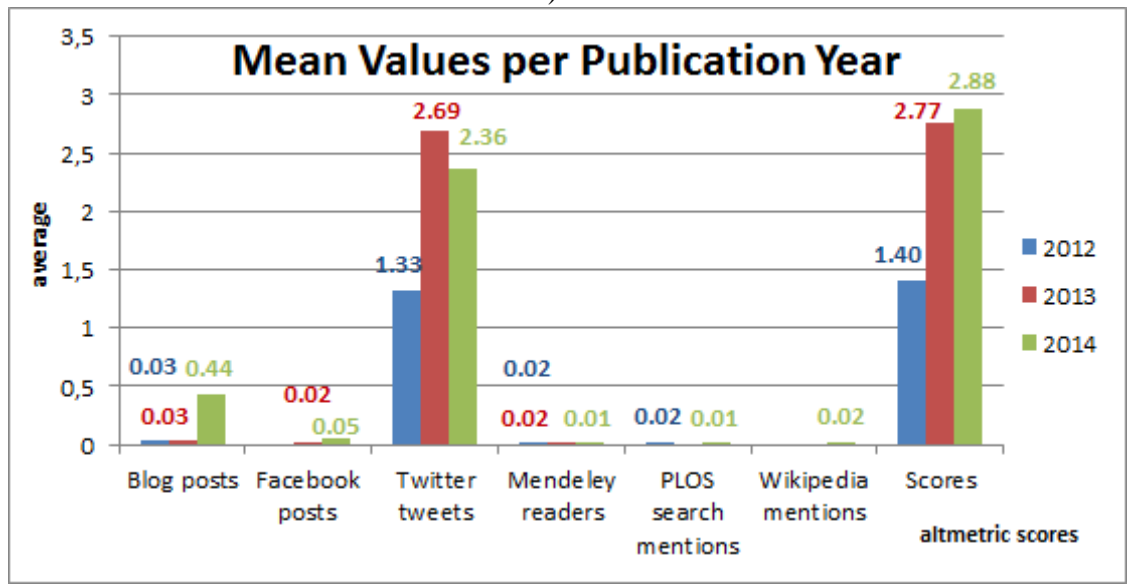

*Comparison of altmetric scores from ImpactStory and PlumX*

Table 8. Results of the comparison of ImpactStory (n=455 items) and PlumX (n=757 items).

| Type | ImpactStory results | | | PlumX results | | |
|---|---|---|---|---|---|---|
| | data | file set | other (incl. article, code, figure, paper, poster, presentation, other) | data | file set | other (incl. article, code, figure, paper, poster, presentation, other) |
| # items | 168 | 278 | 9 | 281 | 456 | 20 |
| # of items with at least one Mendeley-reader | 5 | 2 | 0 | 7 | 14 | 1 |
| # Mendeley-readers total | 6 | 2 | 0 | 8 | 36 | 18 |
| # of items with at least one Wikipedia-mention | 1 | 0 | 0 | 3 | 0 | 0 |
| # Wikipedia-mentions total | 2 | 0 | 0 | 5 | 0 | 0 |
| # of items with at least one Tweet | 44 | 64 | 3 | 176 | 254 | 15 |
| # Tweets total | 365 | 536 | 54 | 2245 | 2638 | 293 |
| # of items with at least one Facebook-post | 3 | 4 | 0 | 281 | 456 | 20 |
| # Facebook-posts total (incl. likes, shares, comments) | 3 | 6 | 0 | 464 | 734 | 26 |
| Total scores | 376 | 544 | 54 | 2722 | 3408 | 337 |
| | ImpactStory results | | | PlumX results | | |

---

[4] Please note: scores for blogs, Facebook and Twitter are provided by altmetric.com for ImpactStory.

Table 8 shows the numbers of items per figshare-document type retrieved from ImpactStory and PlumX as well as the altmetric scores for the social media-platforms both aggregators share. All items (i.e., DOIs) that have been found via ImpactStory (n=455) were also found via PlumX (n=757). PlumX detects considerably more items in social media and also finds higher altmetric scores than ImpactStory. The aggregators, however, differ in the number of social media platforms analysed. For our set of items, ImpactStory uncovers items searched in PLOS and mentioned on blogs. PlumX, on the other hand, reports counts from many other tools not included in ImpactStory (e.g. Reddit; see Tables 5, 6 and 7).

**Discussion and Conclusions**

In our study, we found that almost 90% of all entries in figshare are coming from PLOS. figshare therefore has three basic functions: it acts (1) as a personal repository for yet unpublished materials, (2) as a platform for newly published research materials, and (3) as an archive for PLOS. These different functions are also highlighted by the fact that unviewed and non-downloaded items tend to originate from PLOS. These items are mainly used on the PLOS site, and not on figshare.

It is important to consider the different functions when interpreting the results of a figshare analysis. When analysing the discipline distributions of datasets and filesets, one could easily assume that most users who share their data are from the Natural Sciences (88.9% of all items are assigned to Biological Sciences, Chemistry comes second and Earth Sciences third). More in-depth analysis, however, reveals that the majority of Natural Sciences content is coming from PLOS, and that there seems to be a larger user group sharing datasets coming from Engineering and Social Science. Another unexpected result was that the most shared type of research material is not data, but rather images.

In the altmetrics analysis, we found that Twitter was the social media service where research data gained most attention; generally, research data published in 2014 were most popular across social media services. PlumX detects considerably more items in social media and also finds higher altmetric scores than ImpactStory.

Compared to our previous analysis performed for research data with two or more citations in Data Citation Index (DCI), the following conclusions can be drawn:

- Most research data remain not only uncited but also unviewed/not downloaded.
- Corresponding altmetrics scores for most cited, downloaded and viewed research data are very low, but overall the numbers have been increasing within the last 3 years.
- The results of the comparison of PlumX and ImpactStory are very similar to those obtained in our previous study. In general, comparison of altmetrics tools is difficult due to differences in assignments to categories, which result in different counts. Furthermore, it is hard to judge correctness and completeness of the counts.

**References**

Chamberlain, S. (2013). Consuming article-level metrics: Observations and lessons from comparing aggregator provider data. *Information Standards Quarterly*, 25 (2), 4-13.

figshare (n.d.a). *Which article type?* http://figshare.com/article_types. [Accessed 28 May 2015]

figshare (n.d.b). *FAQs*. Online at http://figshare.com/faqs [Accessed on 17 January 2015]

figshare (2013). *figshare announces partnership with the Public Library of Science*. Online at http://www.digital-science.com/pages/press-releases#figsharePLOS [Accessed on 17 January 2015]

Jobmann, A., Hoffmann, C.P., Künne, S., Peters, I., Schmitz, J., & Wollnik-Korn, G. (2014). Altmetrics for large, multidiciplinary research groups: Comparison of current tools, *Bibliometrie - Praxis und Forschung*, 3. URN:nbn:de:bvb:355-bpf-205-9

Kraker, P., Leony, D., Reinhardt, W., & Beham, G. (2011). The Case for an Open Science in Technology Enhanced Learning. *International Journal of Technology Enhanced Learning*, 6(3), 643–654. doi:10.1504/IJTEL.2011.045454

Kraker, P., Lex, E., Gorraiz, J., Gumpenberger, C., Peters, I. (2015). Figshare Public Metadata until 02/12/2014. figshare. doi:10.6084/m9.figshare.1320834

Peters, I., Kraker, P., Lex, E., Gumpenberger, C., & Gorraiz, J. (2015). Research Data Explored: Citations versus Altmetrics. Digital Libraries. *Submitted to ISSI 2015*. Online at http://arxiv.org/abs/1501.03342 [Accessed 17 January 2015]

Torres-Salinas, D., Martín-Martín, A. & Fuente-Gutiérrez, E. (2013). An introduction to the coverage of the Data Citation Index (Thomson-Reuters): Disciplines, document types and repositories. *EC3 Working Papers*, 11, June 2013. Online at http://arxiv.org/ftp/arxiv/papers/1306/1306.6584.pdf [Accessed on 1 January 2015]

Torres-Salinas, D., Jimenez-Contreras, E. & Robinson-Garcia, N. (2014). How many citations are there in the Data Citation Index? *Proceedings of the STI Conference*, Leiden, The Netherlands. Online at http://arxiv.org/abs/1409.0753 [Accessed on 1 January 2015]

Zahedi, Z., Fenner, M., & Costas, R. (2014). How consistent are altmetrics providers? Study of 1000 PLOS ONE publications using the PLOS ALM, Mendeley and Altmetric.com APIs. In *altmetrics 14*. Workshop at the Web Science Conference, Bloomington, USA. doi:10.6084/m9.figshare.1041821